\documentclass[a4paper]{article}				% Format de page
\usepackage[T1]{fontenc}						% Encodage de caractres
\usepackage{lmodern}						% Police vectorielle Latin Modern
\usepackage[english]{babel}					% Rgles typographiques franaises
\usepackage{amsmath,amsfonts}				% Symboles mathŽmatiques
\usepackage{amssymb}						% Symboles mathŽmatiques bis
\usepackage{graphicx}
\usepackage{epsf, float}
\usepackage{bm}							% Pour les caractres gras en mode mathŽmatiques
\usepackage{amsthm}

\usepackage{a4wide} 						%Pour avoir des marges standard

\newtheorem{thm}{Theorem}[subsection]

\newtheorem{conj}{Conjecture}[subsection]
\newtheorem{pb}{Problem}[subsection]

\theoremstyle{definition}						%Remarques et notations en non italique

\newcommand{\R}{\mathbb{R}}
\newcommand{\C}{\mathbb{C}}
\newcommand{\N}{\mathbb{N}}
\newcommand{\Z}{\mathbb{Z}}
\newcommand{\Q}{\mathbb{Q}}

\title{Open Problems and Conjectures related to the Theory of Mathematical Quasicrystals}
\author{Faustin Adiceam \\ Problem session held in MFO (Germany)}
\date{08 October 2015}

\begin{document}
\maketitle
\tableofcontents

\section{Introduction}

This list of problems arose as a collaborative effort among the participants of the \emph{Arbeitsgemeinschaft} on Mathematical Quasicrystals, which was held at the Mathematisches Forschungsinstitut Oberwolfach in October 2015. The purpose of our meeting was to bring together researchers from a variety of disciplines, with a common goal of understanding different viewpoints and approaches surrounding the theory of mathematical quasicrystals. The problems below reflect this goal and this diversity and we hope that they will motivate further cross-disciplinary research and lead to new advances in our overall vision of this rapidly developing field.\vspace*{.05in}

Most of the technical terms used herein are fairly common in the literature surrounding this subject. When in doubt concerning definitions, the reader may wish to consult \cite{BaakGrim2013} or \cite{Sadu2008}, as well as the references provided with the relevant problem. Here we list some less common terms and abbreviations used below:
\begin{itemize}
\item[(i)] Two point sets in $\R^d$ are {\it bounded displacement equivalent} (abbreviated BD) if there is a bijection between them which moves every point by at most some finite amount.
  \item[(ii)] Two point sets in $\R^d$ are {\it bi-Lipschitz equivalent} (abbreviated BL) if there is a bi-Lipschitz bijection between them.
  \item[(iii)] A pattern $Y\subseteq\R^d$ is {\it linearly repetitive} (abbreviated LR) if there exists a constant $C>0$ such that, for every $r\ge 1$, every patch of size $r$ which occurs anywhere in $Y$, occurs in every ball or radius $Cr$ in $\R^d$.
      \item[(iv)] The abbreviation MLD stands for {\it mutually locally derivable}, as defined in \cite[Section 5.2]{BaakGrim2013}.
\end{itemize}

\section{Problems}

\subsection{D.~Damanik~: Quantum Mechanics and Quasicrystals}

A Schr\"odinger equation associated with a Schr\"odinger operator $H$ can be used to model how well quantum wave packets travel in a quasicrystal --- see~\cite{damaniktchere} for details. A key step in the determination of the large time behaviour of solutions to this equation is the understanding of the properties of the spectrum of $H$ and of the corresponding spectrum measures. Theorem~\ref{thm1} below illustrates the kind of properties that are of interest. Before stating it, we introduce some notation and some definitions.

Given a finite alphabet $\mathcal{A}$, let $\bm{x}\in\mathcal{A}^\N$ (resp.~$\bm{x}\in\mathcal{A}^\Z$) be a single sided (resp.~a double sided) recurrent sequence over  $\mathcal{A}$. The complexity of $\bm{x}$ is defined for each integer $n\ge 1$ as $$p(n)\, :=\, \#\left\{x_m\dots x_{m+n-1}\; : \; m\in\Z \right\}$$ if the sequence is double sided. If it is single sided, the complexity is defined in the same way upon restricting $m$ to the set of positive integers. When $\bm{x}$ is aperiodic (i.e~not ultimately periodic), it is easily seen that $p(n)\ge n+1$ for all $n\ge 1$. The sequence $\bm{x}$ is Sturmian if it is aperiodic and if it has minimal complexity; that is, if  $p(n)= n+1$ for all $n\ge 1$. This definition implies that the alphabet defining a Sturmian sequence contains exactly two letters which may be denoted without loss of generality by 0 and 1. One can then show~\cite{loth} that a double sided (resp.~single sided) Sturmian sequence is exactly of one of the following forms~: for all $m\in\Z$ (resp.~for all $m\in\N$), $$x_m\,:=\, \chi_{[1-\alpha\,;\, 1)}\left(\{m\alpha +\theta\}\right)\qquad \textrm{or}\qquad x_m\,:=\, \chi_{(1-\alpha\,;\, 1]}\left(\{m\alpha +\theta\}\right).$$ Here, $\{x\}$ denotes the fractional part of $x\in\R$, $\chi_E$ the characteristic function of a set $E\subset\R$ and $\alpha$ and $\theta$ are two real numbers such that $\alpha\in\R\backslash\Q$ and $\theta\not\in \alpha\Z+\Z$.

The following result is established in~\cite{bellis} and~\cite{damanikillip}.

\begin{thm}\label{thm1}
Let $\lambda\in\C$ and let $\bm{x}\in\left\{0,1\right\}^\Z$ be a Sturmian sequence. Denote by $H$ the Schr\"odinger operator defined in the space $l^2(\Z)$ as follows~: given $\psi\in l^2(\Z)$ and $m\in\Z$, let $$\left[H\psi\right](m)\,:=\, \psi(m+1)+\psi(m-1)+\lambda\cdot  x_m\cdot \psi(m).$$

Then, the spectrum $\sigma(H)$ of the operator $H$ is a zero Lebesgue measure Cantor set. Furthermore, all spectral measures are singular continuous.
\end{thm}

Another measure of the complexity of a sequence has been introduced by Kamae and Zamboni~\cite{kamzam1, kamzam2}. It is defined as follows (with the natural modification if the sequence is single sided)~: for $n\ge 1$, $$p^*(n)\,:=\, \sup_{0=\tau(0)<\dots<\tau(n-1)}\#\left\{x_{m+\tau(0)}\dots x_{m+\tau(n-1)}\; : \; m\in\Z\right\},$$ where $\tau(0), \dots, \tau(n-1)$ are integers. It is not difficult to see that if $\bm{x}$ is an aperiodic sequence, then $p^*(n)\ge 2n$ for all $n\ge 1$. The sequence $\bm{x}$ is \emph{pattern Sturmian} if $p^*(n)=2n$ for all $n\ge 1$. Note that a Sturmian sequence is necessarily pattern Sturmian. The converse inclusion, however, does not hold.

\begin{pb}
Determine all single sided (resp.~double sided) sequences that are pattern Sturmian.
\end{pb}

With respect to the properties of the Schr\"odinger operator introduced in Theorem~\ref{thm1} above, one can expect the following~:

\begin{conj}
Theorem~\ref{thm1} holds if $\bm{x}$ is a pattern Sturmian sequence.
\end{conj}

\paragraph{} The spectrum of Schr\"odinger operators associated with quasicrystal models in higher dimensions is not well understood. Consider for example the Penrose tiling and the corresponding graph $(V, E)$, where $V$ is the vertex set and $E$ the set of edges. Define the operator $H$ in the space $l^2(V)$ as follows~: for any $\psi\in l^2(V)$ and any $v\in V$, let $$[H\psi](v)\,:=\, \sum_{w\, : \,(v,w)\in E}\left(\psi(w)-\psi(v)\right).$$

\begin{pb}
Determine the spectrum $\sigma(H)$ of the operator defined above.
\end{pb}

\paragraph{} Another example of a two dimensional problem arises when considering a two dimensional potential $V$ that can be written as a sum of two one dimensional potentials $s_1^{\lambda_1}$ and $s_2^{\lambda_2}$. Here, $\lambda_1, \lambda_2\in\C$ and, for $j\in\{1,2\}$ and $k\in\Z$, $$s_j^{\lambda_j}(k) \,:=\, \lambda_j\cdot \chi_{[1-\gamma,1)}(\{k \gamma\})$$ with $\gamma = \frac{\sqrt{5}-1}{2}$. Thus, for any $m, n\in \Z$, $$V(m,n)\, :=\, s_1^{\lambda_1}(m)+s_2^{\lambda_2}(n).$$ The corresponding Schr\"odinger operator $H$ in the space $l^2(\Z^2)$ is defined as follows~: for $\psi\in l^2(\Z)$ and $(m,n)\in\Z^2$, $$[H\psi](m,n)\,:=\, \psi(m+1, n)+\psi(m-1,n)+\psi(m,n+1)+\psi(m,n-1)+V(m,n)\cdot\psi(m,n).$$The spectrum of this operator is expected to exhibit a very particular structure for some values of the parameters $\lambda_1$ and $\lambda_2$~:

\begin{conj}\label{conj01}
There exist values of $\lambda_1$ and $\lambda_2$ such that the spectrum $\sigma(H)$ of $H$ is a Cantorval; that is, the spectrum is the closure of its interior and no connected component is isolated.
\end{conj}

For more details on the concept of Cantorval, see~\cite{cantorval}. With obvious modifications, Conjecture~\ref{conj01} is expected to be true in higher dimensions as well.

\subsection{F.~G{\"a}hler~: The Pisot Substitution Conjecture}

One of the long--standing open problems in the field of mathematical quasicrystals is to determine which tiling dynamical systems (or Delone dynamical systems) have a pure point dynamical spectrum and thus a pure point diffraction pattern. For cut--and--project sets and tilings, this is the case by construction, but for inflation tilings the situation is not as clear. It is known \cite{FG_Sol97} that a self--similar inflation tiling has a non--trivial pure point component in its spectrum if, and only if, the scaling factor of the inflation is a Pisot number $\lambda$; that is, a real algebraic integer $\lambda>1$ all of whose conjugates are strictly smaller than one in modulus. This does not mean,
however, that the spectrum is pure point. There is an algorithm \cite{FG_SS, FG_Sol97} that enables one to check whether a given tiling has pure point spectrum. Nevertheless, simple criteria that are easy to check or known to be true for whole classes of tilings are missing.

The \emph{Pisot Substitution Conjecture} (see also the recent review~\cite{FG_ABBLS}) states the following~:

\begin{conj}[Pisot Substitution Conjecture]
A one--dimensional self--similar inflation tiling with Pisot scaling factor $\lambda$ has pure point spectrum if its abelianisation matrix (i.e.~its substitution matrix) $M$ has an irreducible characteristic polynomial; that is, if the algebraic degree of $\lambda$ equals the dimension of $M$ (or the number of tile types).
\end{conj}

Often, the additional assumption that $M$ is unimodular is made so that $\lambda$ is a unit in the ring $\mathbb{Z}[\lambda]$. So far, this has not really helped to prove the conjecture.

Extensive computer search \cite{FG_AGL} has failed to produce a counter--example but a general proof is also missing. There are some partial results. The conjecture is known to be true in the two tile case ($\lambda$ a quadratic irrational) \cite{FG_HS}. Also, it has been recently been proved for the class of inflation rules which are
injective on the first tile and constant on the last tile \cite{FG_B14} and also for $\beta$--substitutions \cite{FG_B15} (which  do not generally satisfy the conditions of the Pisot Substitution Conjecture).

One problem with the Pisot Substitution Conjecture is that the irreducibility of the characteristic polynomial of $M$ is not invariant under topological conjugacies, whereas the spectral type of the dynamical system is. This has led to the statement of Pisot type conjectures with additional assumptions of a topological nature. The \emph{Homological Pisot Conjecture} \cite{FG_BBJS} is in this vein~:

\begin{conj}[Homological Pisot Conjecture]
A one--dimensional, unimodular Pisot inflation tiling has pure point spectrum if its first rational \v{C}ech cohomology group has rank equal to the algebraic degree of $\lambda$.
\end{conj}

This was later extended to the non--unimodular case in the form of the \emph{Coincidence Rank Conjecture}
\cite{FG_B13}~:

\begin{conj}[Coincidence Rank Conjecture]
The coincidence rank of a one--dimensional Pisot inflation tiling must divide the algebraic norm of $\lambda$.
\end{conj}

The coincidence rank is the multiplicity (almost everywhere) of the factor map to the maximal equicontinuous factor of the tiling dynamical system. It must be one for pure point spectrum.

\subsection{U.~Grimm~: Diffraction of a Pinwheel Tiling}

We first briefly describe the construction of a Pinwheel Tiling following Conway and Radin -- see~\cite{JC, Ra} for further details and~\cite{3} for some of its properties.

Let $\mathcal{T}$ be a right triangle with side lengths 1, 2 and $\sqrt{5}$. As noticed by Conway, $\mathcal{T}$ can be divided into five isometric copies of its image by a dilation of factor $1/\sqrt{5}$ --- see Figure~\ref{fig1} (\footnote{Figures~\ref{fig1}  and~\ref{fig2} are taken from Wikipedia}). A Pinwheel Tiling is then defined as a tiling of the plane whose tiles are isometric copies of $\mathcal{T}$, in which a tile may intersect another tile only either on a whole side or on half the length 2 side, and such that the following property holds~:  the tiles of any Pinwheel Tiling can be grouped in sets of five into homothetic tiles, so that these homothetic tiles form (up to rescaling) a new Pinwheel Tiling.

There are uncountably many Pinwheel Tilings. See Figure~\ref{fig2} below for an example.

\begin{minipage}{0.4\textwidth}
\begin{figure}[H]
\includegraphics[height=3.5cm]{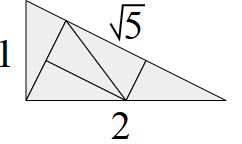}
\caption{{\footnotesize Conway triangle decomposition into homothetic smaller triangles.}}
\label{fig1}
\end{figure}
\end{minipage}
\hspace{8ex}
\begin{minipage}{0.4\textwidth}
\begin{figure}[H]
\includegraphics[height=5.25cm]{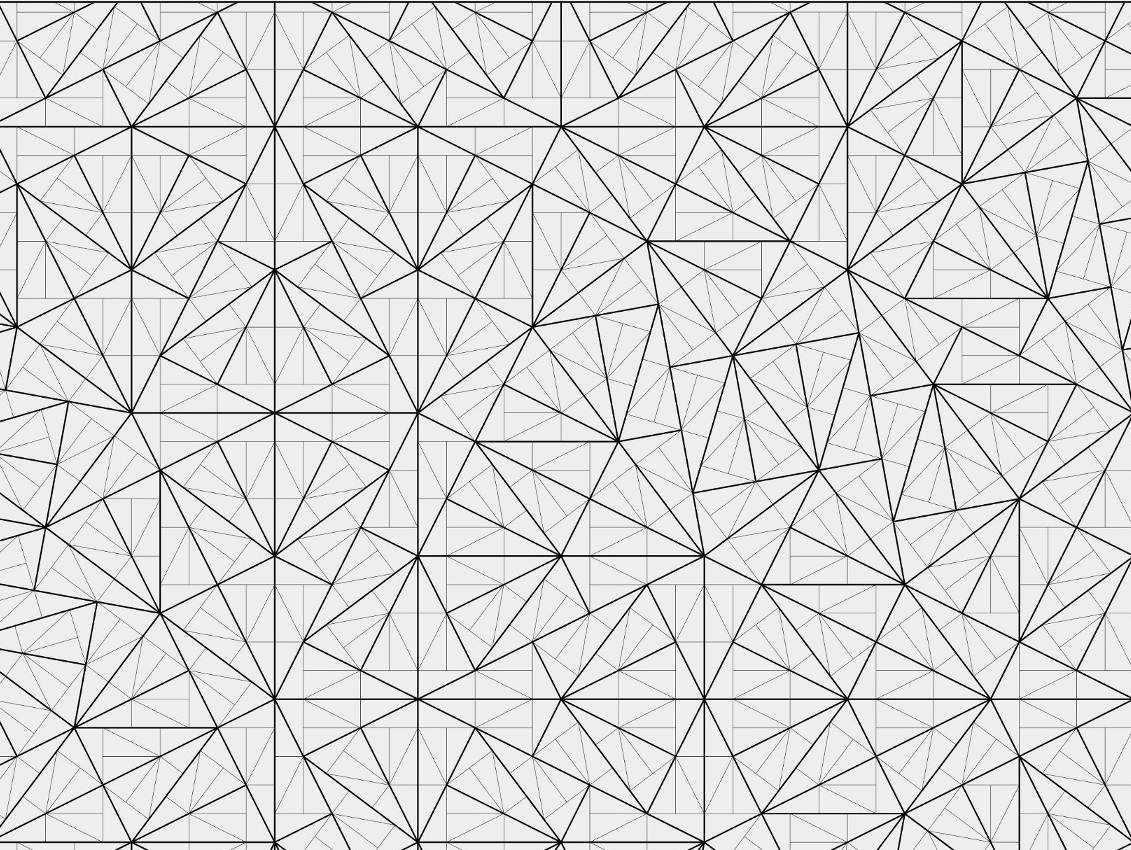}
\caption{{\footnotesize A Pinwheel tiling}}
\label{fig2}
\end{figure}
\end{minipage}

\vspace{3mm}

All Pinwheel Tilings are locally indistinguishable (i.e. any cluster which occurs in one Pinwheel Tiling, occurs in all such tilings) and share the property that tiles appear in infinitely many orientations. Choose a reference point (a natural choice is the point at coordinates $(1/2,\, 1/2)$ with respect to the vertex point at the right--angled corner of the triangular tile) and consider a uniform Dirac comb obtained by placing a point measure at each reference point. The diffraction measure is the Fourier transform of the autocorrelation and is a well--defined positive measure. It is known that this measure has circular symmetry~\cite{4} and, as a consequence, the only pure point part of it is the trivial point measure at the origin, which is related to the density (equal to one in this case). Apart from this point measure, the diffraction measure is continuous. However, note that a measure concentrated on `sharp rings' (so that the measure is `pure point in the radial direction') is a singular continuous measure in the plane as well.

There are arguments that support the existence of such `sharp rings' in the diffraction measure of the pinwheel tiling~\cite{5} similarly to what one would observe for a rotation--averaged square lattice arrangement of point masses. Numerical investigations also indicate the presence of additional components which might be absolutely continuous~\cite{5,6}. However, none of these properties has yet been proved.

\begin{pb}
Determine the position of sharp rings in the diffraction measure of a Pinwheel Tiling and their intensity.
\end{pb}

\begin{pb}
Does the diffraction measure of the Pinwheel Tiling contain an absolutely continuous component?
\end{pb}

\subsection{A.~Haynes~: Gaps Problems}

Let $\alpha, \beta\in\R$ be such that $1, \alpha$ and $\beta$ are $\Q$--linearly independent. Let $Y(\alpha, \beta)$ be a canonical cut--and--project set (this concept is defined in~\cite{hkwcharact}) formed using the subspace $$E(\alpha, \beta)\,:=\, \left\{(x,y, \alpha x + \beta y)\: : \; x,y\in\R \right\} \, \subset \, \R^3.$$

Given a shape $\Omega\subset E(\alpha, \beta)$, let $\xi_{(\alpha, \beta)}(\Omega)$ denote the set of different frequencies  of patches of shape $\Omega$ in $Y(\alpha, \beta)$ which occur in $E(\alpha, \beta)$. Precise definitions of these terms are given in~\cite{hksw} and here we are considering ``type 2 patches''.

Let $M,N\in\N$ and $$S_{(\alpha, \beta)}(M,N)\,:=\, \left\{\{m\alpha+n\beta\}\; : \; 0\le m< M, \, 0\le n< N\right\},$$ where $\{x\}$ denotes the fractional part of $x\in\R$. Depending on the choice of the shape $\Omega$, the cardinality of $\xi_{(\alpha, \beta)}(\Omega)$ is closely related to the number $G_{(\alpha, \beta)}(M,N)$ of distinct lengths of the component intervals of $\mathbb{T}\backslash S_{(\alpha, \beta)}(M,N)$ for specific choices of $M$ and $N$ (here, $\mathbb{T}=\R/\Z$ denotes the one--dimensional torus). For instance, the number of different frequencies for a two--to--one cut--and--project set when the window is an interval is at most 3. This is just another formulation of the Three Distance (or Steinhaus) Theorem and amounts to saying that $G_{(\alpha, \beta)}(1, N)\le 3$ for any $N\ge 1$.

\begin{pb}
Is there a choice of $\alpha$ and $\beta$ as above such that
\begin{equation}\label{lab0}
\sup_{\Omega\in\mathcal{S}} \# \xi_{(\alpha, \beta)}(\Omega)\, = \, +\infty,
\end{equation}where $\mathcal{S}$ denotes the collection of all aligned squares?
\end{pb}

The precise definition of an aligned square (resp.~of an aligned rectangle) can be found in~\cite{hkwperfectly}.

It can be shown~\cite{hksw} that~\eqref{lab0} implies that $\sup_{N\in\N}G_{(\alpha, \beta)}(N, N)\, = \, +\infty$. It was conjectured by Erd\"os~\cite{GS} that the latter equation should hold whenever $1, \alpha$ and $\beta$ are $\Q$--linearly independent. This conjecture was disproved in~\cite{BHJRS}, where it was established that the set of $(\alpha, \beta)$ for which $\sup_{N\in\N}G_{(\alpha, \beta)}(N, N)\, < \, +\infty$, although of zero Lebesgue measure, has full Hausdorff dimension. It is an open problem to determine whether there exists a pair $(\alpha, \beta)$ such that $\sup_{N\in\N}G_{(\alpha, \beta)}(N, N)\, = \, +\infty$.

The following is a seemingly easier version of the previous problem.

\begin{pb}
Is there a choice of $\alpha$ and $\beta$ as above such that
\begin{equation}\label{lab00}
\sup_{\Omega\in\mathcal{R}} \# \xi_{(\alpha, \beta)}(\Omega)\, = \, +\infty,
\end{equation} where $\mathcal{R}$ denotes the collection of aligned rectangles?
\end{pb}

There is currently no known example of a pair $(\alpha, \beta)$ for which~\eqref{lab00} holds. Using the arguments in~\cite{hkwperfectly} and~\cite{hksw}, it can be shown that any such pair satisfies the equation
\begin{equation}\label{lab01}
\liminf_{n\rightarrow\infty} n\left\|n\alpha\right\|\left\|n\beta\right\|  \, = \, 0;
\end{equation}
that is, any such pair satisfies the Littlewood Conjecture (here, $\left\|x\right\|$ denotes the distance from $x\in\R$ to a nearest integer). Whether the converse holds is an open problem~:

\begin{pb}
Given $(\alpha, \beta)\in\R^2$ such that $1, \alpha$ and $\beta$ are $\Q$--linearly independent, does~\eqref{lab01} imply~\eqref{lab00}?
\end{pb}

\subsection{A.~Julien~: Relationship between Complexity and Cohomology}

Consider a tiling and the corresponding tiling space $\Omega$ in dimension $d\ge 1$. In the case of a word $w$ over a finite alphabet (the tiling is then one--dimensional), the complexity function $n\mapsto p(n)$ corresponding to $w$ counts the number of subwords of $w$ of length $n$. In higher dimensions, the complexity function of a tiling counts in an analogous way the number of patches of radius $n$ (up to translation) --- see~\cite{lagpleas} for details.

In the case of a canonical (in particular, irrational and aperiodic) cut--and--project tiling, it is known~\cite[Theorem~5.1]{antoine} that the complexity function grows like $O(n^d)$ if, and only if, the groups of cohomology over $\Q$ of the tiling space are finitely generated. In other words, with usual notation, $$p(n)=O(n^d)\; \iff\; \textrm{rk}(H^*(\Omega, \Q)) \, < \, \infty.$$

The situation is not as well understood in the non--cut--and--project setup. For instance,  there exist tilings in any dimension $d\ge 1$ such that the corresponding tiling spaces have finitely generated groups of cohomology over the rationals whereas their complexity function grows faster than any polynomial -- see, e.g., \cite[Proposition~6.20]{antoine}. In dimension $d=1$ however, the converse can be settled in the affirmative \cite[Proposition~6.7]{antoine}~: if $p(n)=O(n)$, then $\textrm{rk}(H^1(\Omega, \Q)) \, < \, \infty$. The following problem is concerned with a generalization of this result in higher dimensions~:

\begin{pb}
With the same notation, is it the case that for an aperiodic, repetitive tiling of dimension $d$, if $p(n)=O(n^d)$, then $\textrm{rk}(H^*(\Omega, \Q)) \, < \, \infty$?
\end{pb}

Note that the result fails already in dimension one if one considers cohomologies over $\Z$ rather than over $\Q$. Indeed, the substitution defining the Thue--Morse sequence is primitive and therefore~\cite{Queffelec} the corresponding complexity function is bounded above by a linear function. Besides, one can show~\cite{AndersonPutnam} that $$H^1(\Omega_{TM}, \Z)\,\simeq\, \Z\left[\frac{1}{2}\right] \oplus \Z.$$ In particular, $H^1(\Omega_{TM}, \Z)$ is not finitely generated over $\Z$.

\subsection{A.~Navas~: A Conjecture on Delone Sets BL to Lattices (after P.~Alestalo, D.A.~Trotsenko and J.~V\"ais\"al\"a).}

Burago and Kleiner~\cite{burklei} on the one hand and McMullen~\cite{mcmul} on the other proved independently that there exist Delone sets in any dimension which are not BL to any lattice. The following problem, raised in~\cite{additional}, is concerned with the case when a Delone set is BL to a lattice in the plane. As any two lattices are BL, it is enough to consider the case of $\Z^2$.

\begin{pb}\label{conjdelone}
Let $\mathcal{D}\subset\R^2$ be a Delone set BL to $\Z^2$. Does there exist a bi--Lipschitz map $L~:\R^2\mapsto\R^2$ such that $L(\mathcal{D})=\Z^2$?
\end{pb}

In this direction, it has been shown~\cite{nav} that the conclusion of Problem~\ref{conjdelone} is valid under the assumption of linear repetitivity (LR) or under the assumption of Burago-Kleiner (BK). Both of the these conditions imply that a Delone set is BL to a lattice. The reader is referred to~\cite{BK} for an actual definition of the BK condition, which is technical. It is concerned with the speed of convergence of the number of points of a Delone set inside larger and larger balls.

\subsection{L.~Sadun}

\subsubsection{Properties of Patterns and Properties of Tiling Spaces}

There are two very different approaches to tiling theory. A first approach focuses on questions about specific patterns such as~: are they BD or BL to a lattice?  Are they LR? Do they have a pure point diffraction spectrum? Do they have the Meyer property?

A second approach focuses on properties of tiling {\em spaces} such as~: what is the cohomology? What is the  maximal equicontinuous torus? What is the dynamical spectrum?  From a dynamical perspective, tilings that are MLD or that give  topologically conjugate tiling spaces are essentially the same. In some settings, even homeomorphic tilings are considered ``the same''.

Unfortunately, the answers to the first set of questions are not generally preserved inside classes of equivalence of tiling spaces. A natural question is thus to ask whether one can reconcile the two approaches. One strategy to answer this question is to consider a strong version of the BD or BL or LR or $\ldots$ properties.

\begin{pb}
Classify tilings with such a property that any other tiling MLD (or to\-po\-lo\-gi\-cal\-ly conjugate, or homeomorphic) to it also has that property.
\end{pb}

A more ambitious (and open--ended) challenge is the following~:

\begin{pb}
Develop and study new geometric properties, analogous but not identical to BD, BL, etc., that {\em are} invariant under MLD, topological conjugacy, or homeomorphism.
\end{pb}

\subsubsection{Modelling Physical Phenomena with Quasicrystals (after J.~Mi{\k{e}}kisz)}

Many physical phenomena are characterized by short range properties. Examples include thermal stability (temperature is stable under local perturbation), the interactions between two atoms close to each other or else the short range interaction when studying chemical potential. Any quasicrystal model used to describe such phenomena should take into account these local properties.

With this in view, J.~Mi{\k{e}}kisz~\cite{mik1}  labelled two properties that should be satisfied by a local matching rule used to define a tiling~:

\begin{itemize}
\item[(A)] Given a tile $\mathfrak{t}$ and a region $\mathcal{R}$ of the tiling space made of a finite union of tiles, define the discrepancy $D_\mathfrak{t}(\mathcal{R})$ of the tile $\mathfrak{t}$ in $\mathcal{R}$ as $$D_\mathfrak{t}(\mathcal{R})\,:=\, \left|N_\mathfrak{t}(\mathcal{R}) - d(\mathfrak{t})\cdot \mathrm{vol}(\mathcal{R}) \right|.$$ Here, $N_\mathfrak{t}(\mathcal{R})$ stands for the cardinality of the number of tiles $\mathfrak{t}$ in the region $\mathcal{R}$ and $d(\mathfrak{t})$ for the density of the tile $\mathfrak{t}$. The first condition imposed on the tiling is that of ``low fluctuation''; that is, that there exists a constant $c_\mathfrak{t}$ depending only on the tile $\mathfrak{t}$ such that
\begin{equation}\label{lab1}
D_\mathfrak{t}(\mathcal{R})\, \le \, c_\mathfrak{t}\cdot\left|\partial\mathcal{R}\right|,
\end{equation} where $\left|\partial\mathcal{R}\right|$ is the measure of the boundary $\partial\mathcal{R}$ of $\mathcal{R}$. It should be noted that, up to the constant $c_\mathfrak{t}$, one cannot expect a bound better than the right--hand side of~\eqref{lab1} for the discrepancy $D_\mathfrak{t}(\mathcal{R})$ --- see~\cite{mik2} for details.

\item[(B)] The second condition is that property (A) should also hold for any pattern or patch satisfying the given matching rule.
\end{itemize}

\begin{pb}\label{pbphy}
Find a set of matching rules in dimension 2 and/or 3 that meet both condition (A) and condition (B).
\end{pb}

Regarding condition (A), if the system is not uniquely ergodic, then the fluctuations grow like volume (rather than perimeter) and there is no hope to solving the problem. In particular, whenever the problem is relevant, density is well--defined.  Another approach, which does not involve defining density at all, is to look at the difference in population between two different patches of approximately the same volume and to ask whether that difference is bounded by the perimeter(s) and the difference in the volumes. Regarding condition (B), if a patch violates the matching rules at a small set of tiles, one can get another patch that does not violate the rules by simply deleting that small set. This is the main difficulty underlying Problem~\ref{pbphy}.

The following statement can be seen as a first step to the solution of Problem~\ref{pbphy}~:
\begin{pb}\label{matchrule}
Find a set of matching rules in dimension 2 satisfying condition (A) only.
\end{pb}

The contributor is able to produce an example based on the ideas developed in~\cite{mik2} solving the analogue of Problem~\ref{matchrule} in dimension 3. In dimension 2 however, all known examples of matching rules (e.g., that corresponding to the Penrose tiling) have a discrepancy growing like $O\left(\left|\partial\mathcal{R}\right|\cdot \log \left|\partial\mathcal{R}\right|\right)$.

\subsection{B.~Weiss}

\subsubsection{On a Problem of J.~Marklof}

Let $d\ge 1$ be an integer. Denote by $\mathbf{Cl}(\R^d)$ the set of all closed subsets of $\R^d$ with respect to the Chabauty--Fell topology~\cite{harpe}. Thus, $\mathbf{Cl}(\R^d)$ is a compact metric space and any group acting on $\R^d$ acts on it by transporting closed sets.

Let $SL_d(\R)$ denote the group of matrices in dimension $d$ with determinant one, viz.~the group of all volume and orientation preserving linear transformations in $\R^d$. Let $ASL_d(\R)$ denote the affine group in $\R^d$, viz.~the group of all orientation and volume preserving affine maps in $\R^d$.

\begin{pb}[J.~Marklof]
Determine all $SL_d(\R)$--invariant Borel probability measures on $\mathbf{Cl}(\R^d)$ and similarly for the $ASL_d(\R)$ action.
\end{pb}

Examples of such measures include the Dirac masses at $\emptyset$ and at $\R^d$, the measures derived from  a Poisson process, those naturally equipping the space of grids and translated lattices and also those equipping the space of cut--and--project sets.

This problem is motivated by questions in mathematical physics --- see~\cite{marklof1} for some recent work and~\cite{marklof2} for a survey. A topological analogue of this problem was resolved in~\cite{latest}~: it was proved that the only $ASL_d(\R)$--minimal sets are the fixed points $\emptyset, \R^d$ (recall that a minimal set is a closed invariant set with no proper closed invariant subsets).

\subsubsection{A ``Folklore'' Problem concerning the Properties of Cut--and--Project Sets}

In relation with Problem~\ref{conjdelone} above, there is a well--known open problem in the theory of aperiodic tilings which essentially asks whether there exists a cut--and--project set which is not BL to a lattice. The problem can be formulated more rigorously in the following way~:

\begin{pb}
Let $E\subset\R^k$ be a totally irrational subspace of dimension $d\ge 1$, and let $Y$ be a cut--and--project set obtained from $E$ using a bounded window $\mathcal{W}$ with non--empty interior and with the property that the $(k-d)$--dimensional Lebesgue measure of $\partial\mathcal{W}$ is zero. Is such a set $Y$ always BL to a lattice in $E$?
\end{pb}
In \cite{burklei}, it was shown that when $k=3, d=2$ and when $\mathcal{W}$ is an interval, then the set $Y$ is BL to a lattice provided that $E$ satisfies a mild Diophantine condition. In \cite{HKW}, a more general result is proved, which applies for all choices of $k$ and $d$ with a Diophantine hypothesis on $E$, and for all windows $\mathcal{W}$ with the property that the upper Minkowski dimension of $\partial\mathcal{W}$ is less than $k-d$. Further results related to this problem can be found in \cite{H} and \cite{HK}.

It should be noted that if the window in the above problem is only required to be bounded, then it is not difficult (regardless of what $E$ is) to choose $\mathcal{W}$ so that the resulting set $Y$ is not BL to a lattice. To see how to do this, suppose that $k=3$ and $d=2$ and assume without loss of generality that $E+e_3=\R^3$, where $e_3$ denotes a standard basis vector in $\R^3$. Let $F=\R e_3$ and let $\rho_E$ and $\rho_F$ be the projections from $\R^3$ onto $E$ and $F$ with respect to the decomposition $\R^3=E+F$. Let $\mathcal{S}'\subseteq \R^3$ be defined by
\[\mathcal{S}'=\rho_F^{-1}(\{te^3:0\le t<1\}).\]
Start with a Delone set $Y'$ in $\langle e_1,e_2\rangle_\R$ which is a subset of $\langle e_1,e_2\rangle_\Z$, but which is not BL to a lattice in $\langle e_1,e_2\rangle_\R$ (such sets can be produced, for example, by the construction given in \cite{CN}). Then, taking the window to be the bounded set given by
\[\mathcal{W}=\rho_F(\{n\in\Z^3\cap\mathcal{S}':\rho_E(n)\in\rho_{E}(Y')\}),\]
it is clear that the corresponding cut--and--project set $Y$ is not BL to a lattice.

\section{Contact Information}

$\quad$

\begin{minipage}{0.4\textwidth}
Faustin Adiceam\\
Department of Mathematics\\
University of York\\
Heslington\\
York YO10 5DD\\
United Kingdom\\
\texttt{faustin.adiceam@york.ac.uk}\\
\end{minipage}
\hspace{8ex}
\begin{minipage}{0.4\textwidth}
David Damanik\\
Department of Mathematics\\
MS-136, Rice University \\
Houston, TX 77251\\
USA\\
\texttt{damanik@rice.edu}\\
\end{minipage}

\begin{minipage}{0.4\textwidth}
Franz G\"ahler\\
Faculty of Mathematics\\
Bielefeld University\\
D-33615 Bielefeld\\
Germany
\texttt{gaehler@math.uni-bielefeld.de}\\
\end{minipage}
\hspace{8ex}
\begin{minipage}{0.4\textwidth}
Uwe Grimm\\
Department of Mathematics \& Statistics\\
Faculty of Mathematics, Computing \& Technology\\
The Open University \\
Walton Hall\\
Milton Keynes MK7 6AA\\
United Kingdom\\
\texttt{uwe.grimm@open.ac.uk}\\
\end{minipage}

\vspace{2mm}

\begin{minipage}{0.4\textwidth}
Alan Haynes\\
Department of Mathematics\\
University of York\\
Heslington\\
York YO10 5DD\\
United Kingdom\\
\texttt{alan.haynes@york.ac.uk}\\
\end{minipage}
\hspace{8ex}
\begin{minipage}{0.4\textwidth}
Antoine Julien\\
Institut for matematiske fag\\
NTNU \\
7491 Trondheim\\
Norway\\
\texttt{antoine.julien@math.ntnu.no}\\
\end{minipage}

\vspace{2mm}

\begin{minipage}{0.4\textwidth}
Andr\'es Navas\\
Facultad de Ciencia\\
Departamento de Matem\'atica y Ciencia de la Computaci\'on\\
Universidad de Santiago de Chile \\
Chile\\
\texttt{andres.navas@usach.cl}\\
\end{minipage}
\hspace{8ex}
\begin{minipage}{0.4\textwidth}
Lorenzo Sadun\\
Department of Mathematics\\
University of Texas\\
Austin, TX 78712\\
USA\\
\texttt{sadun@math.utexas.edu}\\
\end{minipage}

\begin{minipage}{0.4\textwidth}
Barak Weiss\\
School of Mathematical Sciences\\
Tel Aviv University\\
Israel\\
\texttt{barakw@post.tau.ac.il}\\
\end{minipage}
\hspace{8ex}
\begin{minipage}{0.4\textwidth}
\vspace{30mm}
\end{minipage}

\renewcommand{\abstractname}{Acknowledgements}
\begin{abstract}
The author would like to thank all the contributors (viz.~David Damanik, Franz G\"ahler, Uwe Grimm, Alan Haynes, Antoine Julien, Andr\'es Navas, Lorenzo Sadun \& Barak Weiss) for providing such interesting problems and for their careful reading of a preliminary draft of the paper. He would in particular like to thank Prof.~Barak Weiss for suggesting the idea of this problem set and for his help during its elaboration. The introduction was written up by Dr.~Alan Haynes.
The author's work was supported by EPSRC Programme Grant EP/J018260/1.
\end{abstract}


\begin{thebibliography}{99}

\bibitem{BaakGrim2013}
M.~Baake and U.~Grimm.
\emph{Aperiodic order. Vol. 1. A mathematical invitation},
Encyclopedia of Mathematics and its Applications, 149, Cambridge University Press, Cambridge,  2013.

\bibitem{Sadu2008}
L.~ Sadun.
\emph{Topology of tiling spaces}, University Lecture Series, 46, American Mathematical Society, Providence, RI,  2008.
\end{thebibliography}

\begin{thebibliography}{99}

\bibitem{bellis} J.~Bellissard, B.~Iochum, E.~Scoppola and D.~Testard,
\newblock Spectral properties of one-dimensional quasicrystals,
\newblock{\em Comm.~Math.~Phys.}~125, 527--543 (1989).

\bibitem{damanikillip} D.~Damanik, R.~Killip, and D.~Lenz,
\newblock Uniform spectral properties of one-dimensional quasicrystals. III. $\alpha$--continuity,
\newblock{\em Comm.~Math.~Phys.}~212(1), 191--204 (2000).

\bibitem{damaniktchere} D.~Damanik and S.~Tcheremchantsev,
\newblock A general description of quantum dynamical spreading over an orthonormal basis and applications to Schr\"odinger operators,
\newblock{\em Discrete Contin.~Dyn.~Syst.}~A 28, 1381--1412 (2010).

\bibitem{kamzam1} T.~Kamae and L.~Zamboni,
\newblock Maximal pattern complexity for discrete systems,
\newblock{\em Ergodic Theory Dynam.~Systems}~22(4), 1201--1214 (2002).

\bibitem{kamzam2} T.~Kamae and L.~Zamboni,
\newblock Sequence entropy and the maximal pattern complexity of infinite words,
\newblock{\em Ergodic Theory Dynam.~Systems}~22(4), 1191--1199 (2002).

\bibitem{loth} M.~Lothaire,
\newblock {\em Algebraic combinatorics on words. Encyclopedia of Mathematics and its Applications}.
\newblock Encyclopedia of Mathematics and its Applications, 90. Cambridge University Press, Cambridge, 2002.

\bibitem{cantorval} P.~Mendes and F.~Oliveira,
\newblock On the topological structure of the arithmetic sum of two Cantor sets,
\newblock{\em Nonlinearity}~7(2), 329--343 (1994).

\end{thebibliography}

\begin{thebibliography}{99}

\bibitem{FG_ABBLS}
S.~Akiyama, M.~Barge, V.~Berth\'e, J.-L.~Lee, and A.~Siegel,
\textit{On the Pisot Substitution Conjecture},
in \textit{Mathematics of Aperiodic Order}, eds.\ J.~Kellendonk, D.~Lenz,
and J.~Savinien, Progress in Mathematics 309, 33--72  (2015).

\bibitem{FG_AGL}
S.~Akiyama, F.~G\"ahler and J.-L.~Lee,
\textit{Determining pure discrete spectrum for some self-affine tilings},
DMTCS 16(3), 305--316 (2014).

\bibitem{FG_B13}
M.~Barge, \textit{Factors of Pisot tiling spaces and the Coincidence Rank
Conjecture}, Bull.\ Soc.\ Math.\ France 143, 357--381 (2015).

\bibitem{FG_B15}
M.~Barge, \textit{The Pisot conjecture for $\beta$-substitutions},
preprint, \texttt{arXiv:1505.04408} (2015).

\bibitem{FG_B14}
M.~Barge, \textit{Pure discrete spectrum for a class on one-dimensional
substitution tiling systems},
Disc.\ Cont.\ Dynam.\ Sys.\ A 36, 1159--1173 (2016).

\bibitem{FG_BBJS}
M.~Barge, H.~Bruin, L.~Jones, and L.~Sadun,
\textit{Homological Pisot substitutions and exact regularity},
Israel J.\ Math.\ 188, 281--300 (2012).

\bibitem{FG_HS}
M.~Hollander and B.~Solomyak,
\textit{Two-symbol Pisot substitutions have pure discrete spectrum},
Ergod.\ Th.\ \& Dynam.\ Syst.\ 23, 533--640 (2003).

\bibitem{FG_SS}
V.F.~Sirvent and B.~Solomyak,
\textit{Pure discrete spectrum for one-dimensional substitution
systems of Pisot type},
Canad.\ Math.\ Bull.\ 45, 697--710 (2002).

\bibitem{FG_Sol97}
B.~Solomyak, \textit{Dynamics of tiling spaces},
Ergod.\ Th.\ \& Dynam.\ Syst.\ 17, 695--738 (1997).

\end{thebibliography}

\begin{thebibliography}{99}

\bibitem{5} M.~Baake, D.~Frettl\"{o}h and U.~Grimm,
\newblock A radial analogue of Poisson's summation formula with applications to powder diffraction and pinwheel patternsn,
\newblock{\em J.\ Geom.\ Phys.}~57, 1331--1343 (2007).

\bibitem{JC} J.~Conway, C.~Radin,
\newblock Quaquaversal tilings and rotations,
\newblock{\em Inv.~Math}.~132, 179--188 (1998).

\bibitem{6} U.~Grimm and X.~Deng,
\newblock Some comments on pinwheel tilings and their diffraction,
\newblock{\em J.\ Phys.: Conf.\ Ser.}~284, 012032 (2011).

\bibitem{3} H.~Moustafa,
\newblock PV cohomology of the pinwheel tilings, their integer group of coinvariants and gap-labeling,
\newblock{\em Commun.\ Math.\ Phys.}~298, 369--405 (2010).

\bibitem{4} R.V.~Moody, D.~Postnikoff and N.~Strungaru,
\newblock Circular symmetry of pinwheel diffraction,
\newblock{\em Ann.\ Henri Poincar\'{e}}~7, 711--730 (2006).

\bibitem{Ra} C.~Radin,
\newblock The Pinwheel Tilings of the Plane,
\newblock{\em Ann Math}.~139(3), 661--702 (1994).

\end{thebibliography}

\begin{thebibliography}{99}

\bibitem{BHJRS} P.M.~{Bleher}, Y~{Homma}, L.L.~{Ji}, R.K.W.~{Roeder} and J.D.~{Shen},
\newblock Nearest neighbor distances on a circle~: Multidimensional case,
\newblock{\em J.~Stat.~Phys}.~146(2), 446--465 (2012).

\bibitem{GS} J.F.~Geelen and R.J.~Simpson,
\newblock A two dimensional Steinhaus theorem,
\newblock{\em Australas.~J.~Combin}.~8, 169--197 (1993).

\bibitem{hkwcharact} A.~Haynes, H.~Koivusalo and J.~Walton,
\newblock A characterization of linearly repetitive cut and project sets,
\newblock{\em submitted}, available at~\texttt{arxiv:1503.04091}.

\bibitem{hkwperfectly} A.~Haynes, H.~Koivusalo and J.~Walton,
\newblock Perfectly ordered quasicrystals and the Littlewood Conjecture,
\newblock{\em submitted}, available at~\texttt{arxiv:1506.05649}.

\bibitem{hksw} A.~Haynes, H.~Koivusalo, L.~Sadun and J.~Walton,
\newblock Gaps problems and frequencies of patches in cut and project sets,
\newblock{\em submitted}, available at~\texttt{arxiv:1411.0578}.


\end{thebibliography}

\begin{thebibliography}{99}

\bibitem{AndersonPutnam} J.E.~Anderson and I.F.~Putnam,
\newblock Topological invariants for substitution tilings and their associated $C\sp *$-algebras,
\newblock{\em Ergodic Theory Dynam.~Systems} 18(3), 50--537 (1998).

\bibitem{antoine} A.~Julien,
\newblock Complexity and cohomology for cut--and--projection tilings,
\newblock{\em Ergodic Theory Dynam.~Systems} 30(2), 48--523 (2010).

\bibitem{lagpleas} J.C.~Lagarias and P.A.B~Pleasants,
\newblock Repetitive Delone sets and quasicrystals,
\newblock{\em Ergodic Theory Dynam.~Systems} 23(3), 83--867 (2003).

\bibitem{Queffelec} M.~Queff\'elec,
\newblock {\em Substitution dynamical systems--spectral analysis. Second edition.}
\newblock  Lecture Notes in Mathematics, 1294. Springer--Verlag, Berlin, 2010.


\end{thebibliography}

\begin{thebibliography}{99}


\bibitem{additional} P.~Alestalo, D.A.~Trotsenko and J.~V\"ais\"al\"a,
\newblock Linear Bilipschitz Extension Property,
\newblock{\em Sibirsk. Mat. Zh.} 44(6), 1226--1238 (1993). Translation into English in {\em Siberian Mathematical
Journal} 44(6), 959--968 (1993).

\bibitem{BK} D.~Burago and B.~Kleiner,
\newblock Separated nets in Euclidean space and Jacobians of bi--Lipschitz maps,
\newblock{\em Geom.~Funct.~Anal}.~8(2), 273--282 (1998).

\bibitem{burklei} D.~Burago and B.~Kleiner,
\newblock Rectifying separated nets,
\newblock{\em Geom.~Funct.~Anal}.~12(1), 80--92 (2002).

\bibitem{mcmul} C.T.~McMullen,
\newblock Lipschitz maps and nets in Euclidean space,
\newblock{\em Geom.~Funct.~Anal}.~8(2), 304--314 (1998).

\bibitem{nav} A.~Navas,
\newblock A remark concerning bi-Lipschitz equivalence of Delone sets,
\newblock notes available from the author (not intended for publication).

\end{thebibliography}

\begin{thebibliography}{99}

\bibitem{mik1} J.~Mi{\k{e}}kisz,
\newblock An ultimate frustration in classical lattice-gas models.
\newblock{\em J.~Stat.~Phys.}~90, 285--300 (1998).

\bibitem{mik2} J.~Mi{\k{e}}kisz,
\newblock Classical lattice--gas models of quasicrystals,
\newblock{\em J.~Stat.~Phys.}~95, 835--850 (1999).

\end{thebibliography}

\begin{thebibliography}{99}

\bibitem{harpe} P.~de la Harpe,
\newblock Spaces of closed subgroups of locally compact groups.
\newblock Arxiv:0807.2030.

\bibitem{marklof2} J.~Marklof,
\newblock {\em Kinetic limits of dynamical systems},
\newblock in: Hyperbolic Dynamics, Fluctuations and Large Deviations, eds. D. Dolgopyat, Y. Pesin, M. Pollicott, L. Stoyanov, Proc. Symp. Pure Math., American Mathematical Soc., 195--223 (2015).

\bibitem{marklof1} J.~Marklof and A.~Str\"ombergsson,
\newblock Free path lengths in quasicrystals.
\newblock{\em Comm.  Math. Phys.}~330, 723--755 (2014) .

\bibitem{latest} O.~Solan, Y~Solomon and B.~Weiss,
\newblock On problems of Danzer and Gowers and dynamics on the space of closed subsets of $R^d$.
\newblock Preprint (2015).



\end{thebibliography}

\begin{thebibliography}{99}

\bibitem{burklei} D.~Burago and B.~Kleiner,
\newblock Rectifying separated nets,
\newblock{\em Geom.~Funct.~Anal}.~12(1), 80--92 (2002).

\bibitem{CN}
M.~I.~Cortez and A.~Navas,
\newblock Some examples of non--rectifiable, repetitive Delone sets,
\newblock {\em preprint}, available at~\texttt{arxiv:1401.7927}.

\bibitem{H} A.~Haynes,
\newblock Equivalence classes of codimension one cut--and--project nets,
\newblock{\em Ergodic Theory Dyn.~Syst.}, to appear.

\bibitem{HKW} A.~Haynes, M.~Kelly and B.~Weiss,
\newblock Equivalence relations on separated nets arising from linear toral flows,
\newblock{\em Proc.~Lond.~Math.~Soc.~(3)}, 109(5), 1203--1228 (2014).

\bibitem{HK} A.~Haynes and H.~Koivusalo,
\newblock Constructing bounded remainder sets and cut--and--project sets which are bounded distance to lattices,
\newblock{\em Israel J.~Math.}, to appear.

\end{thebibliography}
\end{document}